\documentclass[journal,twoside,web]{ieeecolor}
\usepackage{generic}
\usepackage{cite}
\usepackage{amsmath,amssymb,amsfonts}
\usepackage{algorithmic}
\usepackage{graphicx}
\usepackage{algorithm,algorithmic}
\usepackage{hyperref}
\usepackage{textcomp}
\usepackage{bm}

\newtheorem{theorem}{Theorem}
\newtheorem{proposition}{Proposition}
\newtheorem{lemma}{Lemma}
\newtheorem{corollary}{Corollary}
\newtheorem{remark}{Remark}

\newtheorem{assumption}{Assumption}

\newcommand{\revision}[1]{\textcolor{black}{#1}}

\begin{document}
\title{B-Spline Neural Operators for Nonlinear Dynamical Systems: Representation and Analysis}
\author{Raffaele Romagnoli, \IEEEmembership{Member, IEEE}, Soummya Kar, \IEEEmembership{Fellow, IEEE}
\thanks{Romagnoli R. is with the Department of Mathematics and Computer Science, Pittsburgh, PA 15282 USA (e-mail: romagnolir@duq.edu). }
\thanks{Kar S. is with the Department of Electrical and Computer Engineering, Carnegie Mellon University,  Pittsburgh, PA 15213 USA (e-mail: soummyak@andrew.cmu.edu).}}

\maketitle

\begin{abstract}
B-spline Neural Operators (BNOs) approximate continuous-time solution trajectories through a finite set of B-spline control points. This paper develops a dynamical analysis framework that
operates directly on this native representation, without reconstructing and sampling the predicted trajectory. We establish a quantitative relation between the control points and the underlying continuous-time dynamics and show how the
resulting control-point sequence can be analyzed using Dynamic Mode Decomposition (DMD) and Hankel Dynamic Mode Decomposition (HDMD). We further establish conditions relating convergence of the control-point representation to equilibrium convergence of the underlying system and develop a finite-horizon spectral framework for stability assessment. Numerical experiments on linear and nonlinear systems demonstrate the proposed framework across stable, periodic, and chaotic dynamical regimes.
\end{abstract}


\section{Introduction}
\label{sec:introduction}
The rapid development of Scientific Machine Learning (SciML) has led to a new generation of learning frameworks capable of approximating complex nonlinear dynamical systems directly from data~\cite{dietrich2025scientific, azizzadenesheli2024neural}.
Among these approaches, universal Neural Operators (NOs), such as DeepONets~\cite{lu2021learning} and Fourier Neural Operators (FNOs)~\cite{li2020fourier}, have emerged as powerful surrogate models that learn mappings between function spaces and generalize across discretizations, initial conditions, and parameter variations. Complementary approaches~\cite{mezic2021koopman, brunton2021modern, mauroy2020koopman} instead seek to exploit known mathematical structures, such as the Koopman operator, to learn interpretable representations of nonlinear dynamics directly from data. These developments have significantly expanded the role of data-driven models in engineering, scientific computing, and increasingly in control applications~\cite{azizzadenesheli2024neural, brunton2022data}.

While the predictive capabilities of neural operators have advanced remarkably, considerably less attention has been devoted to what can be done once the training process is completed. In many engineering applications, obtaining an accurate surrogate model is only the first step. Equally important is understanding the dynamical information encoded by the learned representation.

Recent work has begun to explore the information contained in trained neural operators beyond prediction accuracy. Examples include uncertainty quantification~\cite{weber2024uncertainty}, regularization techniques for stable latent dynamics~\cite{guo2024surprising}, and, more recently, Equation-Free methodologies that use trained neural operators for stability and bifurcation computations~\cite{fabiani2026enabling}. These developments indicate a growing interest in analyzing the properties of learned models beyond predictive accuracy within SciML.

Despite these efforts, establishing a direct connection between the finite-dimensional representation produced by a trained neural operator and the dynamics of the underlying continuous-time system remains nontrivial. Such a connection is particularly relevant when the learned representation is used beyond trajectory prediction, for example for stability and spectral analysis. These properties play a central role in learning-enabled control, runtime assurance, and predictive safety filtering~\cite{hewing2020learning,wabersich2021predictive,ames2019control}, where guarantees on stability and safety require information about the underlying dynamics rather than prediction accuracy alone.

One approach is to incorporate dynamical structure directly during training. Koopman-based learning methods~\cite{lusch2018deep,bevanda2021koopman,mitjans2024learning} seek latent coordinates that evolve approximately linearly, thereby enabling spectral analysis through finite-dimensional approximations of the Koopman operator. While this philosophy provides an elegant framework for dynamical analysis, it explicitly constructs or constrains the learned representation according to a prescribed dynamical structure. In contrast, this work considers the post-training analysis of a neural operator whose representation is determined by its original approximation task, without modifying the architecture or imposing additional dynamical constraints during training.

This paper develops this approach by considering B-spline Neural
Operators (BNOs)~\cite{romagnoli_hybrid, wang2025physics}.
\revision{BNOs learn a continuous-time solution operator by mapping an initial condition to a finite set of B-spline control points defining the predicted trajectory. This provides a reusable surrogate when the governing model is unavailable or when acquiring additional trajectories from prescribed initial conditions is impractical. Unlike pointwise trajectory samples, the control points are coefficients of the continuous-time B-spline representation and constitute the native finite-dimensional output of the trained operator.}

\revision{The objective of this paper is to establish a dynamical interpretation of this control-point representation and use it directly for spectral and stability analysis. We derive approximation bounds relating the control points to the continuous-time trajectory and establish a bounded dynamical relation between consecutive control points and the sampled
continuous-time flow. This connection enables Dynamic Mode Decomposition
(DMD)~\cite{schmid2010dynamic,tu2014dynamic} and Hankel Dynamic Mode Decomposition (HDMD)~\cite{arbabi2017ergodic} to be applied directly to the control-point sequence, without first evaluating the learned continuous-time trajectory on an additional temporal sampling grid. Importantly, the control points are not constructed as Koopman observables; they arise directly from the original trajectory approximation task.}

The main contributions of this paper are summarized as follows.
\begin{itemize}

\item
We establish a dynamical interpretation of the native control-point
representation generated by trained BNOs.

\item
We derive approximation bounds relating the control points to the
continuous-time trajectory and establish a bounded dynamical relation
between the ordered control-point sequence and the sampled
continuous-time flow.

\item
We relate convergence of the control-point representation to
equilibrium convergence of the underlying continuous-time system and
develop a finite-horizon spectral framework based on DMD and HDMD,
together with a finite-data richness condition for the identification
data.

\item
We evaluate the framework on linear and nonlinear systems with stable
equilibria, periodic orbits, and chaotic dynamics, illustrating the
distinction between equilibrium-convergence assessment and
finite-horizon spectral characterization.

\end{itemize}

\section{Preliminaries}
\label{sec:preliminaries}
\subsection{Continuous-Time Nonlinear Systems}

Consider the autonomous nonlinear system

\begin{equation}
\dot{x}(t)=f(x(t)),
\label{eq:cts}
\end{equation}

where $x(t)\in D\subseteq\mathbb{R}^{n}$ and
$f:D\rightarrow\mathbb{R}^{n}$ is continuously
differentiable. \revision{The vector field $f$ is assumed to be unknown, and the proposed framework does not require access to an analytical model of the system.} Throughout this paper, it is assumed that solutions exist for all $t\ge0$ and remain in a
compact forward-invariant set $K\subseteq D$~\cite{Khalil:1173048}. The solution admits the integral representation

\begin{equation}
x(t;x_0)
=
x_0
+
\int_{0}^{t}
f(x(\tau;x_0))\,d\tau .
\label{eq:integral_solution}
\end{equation}

The flow of \eqref{eq:cts} induces the solution operator $\mathcal{P}:K\rightarrow C([0,\infty),\mathbb{R}^n)$,
defined by
\begin{equation}
    \mathcal{P}(x_0)
    =
    x(\cdot;x_0),
    \label{eq:solution_operator}
\end{equation}
which maps each initial condition to its corresponding trajectory.
Denoting by $\Phi_t:D\rightarrow D$ the flow generated by
\eqref{eq:cts}, $x(t;x_0)=\Phi_t(x_0)$. For sampling instants
$\{t_k\}_{k\geq0}$, let $x_k:=x(t_k)=\Phi_{t_k}(x_0)$.

The sampled dynamics satisfy
\begin{equation}
x_{k+1}=\Phi_{\Delta_k}(x_k),
\qquad
\Delta_k=t_{k+1}-t_k,
\label{eq:sampled_dynamics_variable}
\end{equation}
which reduces to uniform sampling when $\Delta_k=\Delta$.

Since $f$ is continuously differentiable, the solution is twice
continuously differentiable on every finite interval, with
$\ddot{x}(t)=Df(x(t))f(x(t))$. This regularity is used in
Section~V to characterize the control-point approximation error.

\subsection{B-Spline Representation}

The trajectory $x(t)$ is approximated by the vector-valued spline $s:I\rightarrow\mathbb{R}^n$, where each state component admits the scalar B-spline expansion

\begin{equation}
s_r(t)
=
\sum_{i=1}^{\ell}
c_{i,r} B_{i,d}(t),
\qquad
r=1,\ldots,n,
\label{eq:scalar_spline}
\end{equation}

where $B_{i,d}(t)$ denotes the $i$-th B-spline basis
function of degree $d$, which is defined
recursively by the Cox--de~Boor \cite{de1978practical, schumaker2007spline} formula

\begin{equation}\label{cox_de_boor}
B_{i,d}(t)
=
\frac{t-\tau_i}{\tau_{i+d}-\tau_i}\,
B_{i,d-1}(t)
+
\frac{\tau_{i+d+1}-t}{\tau_{i+d+1}-\tau_{i+1}}\,
B_{i+1,d-1}(t),
\end{equation}

with the zeroth-degree basis functions

\[
B_{i,0}(t)
=
\begin{cases}
1,
&
\tau_i\le t<\tau_{i+1},
\\[2mm]
0,
&
\text{otherwise}.
\end{cases}
\]

The B-spline coefficient $c_{i,r}\in\mathbb{R}$ is the corresponding scalar control point. All state components share the same knot vector

\begin{equation}
\tau=
\{\tau_1,\tau_2,\ldots,\tau_{\ell+d+1}\},
\label{eq:knot_vector}
\end{equation}
with $\tau_1\leq\tau_2\leq\cdots\leq\tau_{\ell+d+1},$ and $I=[\tau_1,\tau_{\ell+d+1}]$. The corresponding scalar spline space is
$S_{d,\tau}=\operatorname{span}\{B_{i,d}\}_{i=1}^{\ell}$,
so that each $s_r$ in \eqref{eq:scalar_spline} belongs to
$S_{d,\tau}$.

Since the same basis functions are used for all state components, the scalar coefficients associated with the same basis function are grouped into the vector-valued control point

\begin{equation}
c_i =
\begin{bmatrix}
c_{i,1}\\
c_{i,2}\\
\vdots\\
c_{i,n}
\end{bmatrix}
\in\mathbb{R}^n,
\qquad
i=1,\ldots,\ell.
\label{eq:vector_control_point}
\end{equation}

Using \eqref{eq:vector_control_point}, the vector-valued spline representation can be written compactly as

\begin{equation}
s(t)
=
\sum_{i=1}^{\ell}
B_{i,d}(t)\,c_i,
\label{eq:vector_spline}
\end{equation}

where $s(t)\in\mathbb{R}^n$.

The knot averages, also known as Greville abscissae \cite{prautzsch2002bezier}, are defined as

\begin{equation}
\tau_i^*
=
\frac{\tau_{i+1}+\tau_{i+2}+\cdots+\tau_{i+d}}{d},
\qquad
i=1,\ldots,\ell.
\label{eq:greville}
\end{equation}

Each control point $c_i$ is associated with the
corresponding Greville abscissa $\tau_i^*$ and, as shown
below, approximates the value of the corresponding spline
at $\tau_i^*$.

An open (clamped) knot vector is adopted by repeating
the first and last $d+1$ knots,

\begin{equation}
\tau_1=\cdots=\tau_{d+1},
\qquad
\tau_{\ell+1}=\cdots=\tau_{\ell+d+1}.
\label{eq:open_knot}
\end{equation}

This guarantees endpoint interpolation,
$s(\tau_1)=c_1$ and $s(\tau_{\ell+d+1})=c_\ell$.
Since the initial condition is prescribed, $c_1=x_0$. The last control point coincides with the terminal value
of the represented trajectory on the interval
$I=[\tau_{d+1},\tau_{\ell+d+1}]$.

Throughout, we use
$\|c\|_\infty:=\max_{1\leq i\leq\ell}\|c_i\|$ and
$\|v\|_{L^\infty(I)}:=\sup_{t\in I}\|v(t)\|$, where
$\|\cdot\|$ denotes a fixed norm on $\mathbb{R}^n$.

The following result relates the control points to the values of the
spline~\cite{de1978practical,lyche2008spline}.

\begin{lemma}
\label{lem:control_point_proximity}
Let $s$ be the vector-valued spline
\eqref{eq:vector_spline}. Then
\revision{
\[
\|c_i-s(\tau_i^*)\|
\leq
C_d
(\tau_{i+d}-\tau_{i+1})^2
\|D^2s\|_{L^\infty([\tau_{i+1},\tau_{i+d}])},
\]
}
for $i=1,\ldots,\ell$, where
\revision{$C_d>0$ is a constant depending only on the spline degree $d$.}
\end{lemma}

The previous result implies that, provided the spline
curvature remains uniformly bounded under knot refinement,
the control points converge quadratically to the values
of the spline evaluated at the corresponding Greville
abscissae as the knot spacing tends to zero.

\subsection{Koopman Operator Theory and Data-Driven Approximations}

Consider the discrete-time nonlinear system
\begin{equation}
    x_{k+1}=F(x_k),
    \label{eq:discrete_nonlinear_system}
\end{equation}
where $F:D\rightarrow D$ denotes the discrete-time evolution map. The Koopman operator provides an alternative description of the nonlinear dynamics through the evolution of observable quantities. For a scalar observable $\varphi:D\rightarrow\mathbb C$, the
Koopman operator $\mathcal U$ is defined by
\begin{equation}
(\mathcal U\varphi)(x)=\varphi(F(x)).
\label{eq:koopman_operator}
\end{equation}
Although $F$ may be nonlinear, $\mathcal U$ is linear on the
generally infinite-dimensional space of observables.

For a finite collection of observables
$g(x)=[g_1(x),\ldots,g_m(x)]^\top$, let $z_k=g(x_k)$.
If their span is Koopman invariant, the observable coordinates
admit an exact finite-dimensional evolution $z_{k+1}=Az_k$;
otherwise, one seeks the approximation
\begin{equation}
z_{k+1}\approx Az_k.
\label{eq:finite_koopman_approx}
\end{equation}
DMD identifies such a matrix directly from snapshot pairs
$(z_k,z_{k+1})$~\cite{tu2014dynamic}. When $g(x)=x$, DMD
operates directly on sampled state trajectories.

HDMD augments these coordinates with time delays,
\begin{equation}
\zeta_k =
\begin{bmatrix}
z_k^\top & z_{k-1}^\top & \cdots & z_{k-q+1}^\top
\end{bmatrix}^{\!\top},
\label{eq:delay_coordinates_prelim}
\end{equation}
and applies DMD to the sequence $\{\zeta_k\}$. Neither DMD nor
HDMD requires the selected coordinates to span an exact
finite-dimensional Koopman-invariant subspace; the resulting operator
instead approximates their observed sequential evolution
\cite{tu2014dynamic}. This motivates investigating the BNO
control-point representation directly as coordinates for spectral
analysis.

\section{B-Spline Neural Operator Representation}

This section introduces a general class of neural
operators whose outputs are represented through
B-spline expansions. Such representations arise in
several operator-learning architectures and provide a
finite-dimensional parameterization of continuous-time
trajectories through a sequence of control points. The
subsequent analysis is developed at the level of the
B-spline representation and is therefore independent of
the particular neural network architecture used to
generate the control points.

Consider the solution operator
$\mathcal{P}$ introduced in Section~II-A. \revision{A BNO is trained from trajectory data to approximate
$\mathcal{P}$ over a compact set
$K\subseteq D$ of initial conditions, without requiring
explicit knowledge of the vector field $f$}. The learned
operator

\[
\widehat{\mathcal{P}}
:
K
\rightarrow
C([0,T],\mathbb{R}^n)
\]

maps an initial condition $x_0$ to an approximation of
the corresponding solution trajectory of
\eqref{eq:cts} over the finite horizon $[0,T]$.

A B-spline neural operator combines a neural network
with the B-spline representation
\eqref{eq:vector_spline}. Given an initial condition
$x_0$, the neural network generates the vector-valued
control-point sequence

\[
c(x_0)
=
\begin{bmatrix}
c_1(x_0)\\
c_2(x_0)\\
\vdots\\
c_\ell(x_0)
\end{bmatrix},
\qquad
c_i(x_0)\in\mathbb{R}^n,
\]

which parameterizes the predicted trajectory through
the spline expansion

\begin{equation}
\hat{x}(t;x_0)
=
\sum_{i=1}^{\ell}
B_{i,d}(t)\,c_i(x_0).
\label{eq:hbdno_output}
\end{equation}
\revision{Thus, the control-point sequence is the native
finite-dimensional representation generated by the BNO,
while the continuous-time prediction is obtained by
evaluating the corresponding B-spline expansion.}
Equivalently,

\begin{equation}
\hat{x}(t;x_0)
=
\widehat{\mathcal{P}}(x_0)(t),
\label{eq:hbdno_operator}
\end{equation}

where $\widehat{\mathcal{P}}$ denotes the neural
operator induced by the network and the B-spline basis.

By the open knot construction
\eqref{eq:open_knot}, the first control point is
constrained to satisfy

\begin{equation}
c_1(x_0)=x_0,
\label{eq:hbdno_initial_condition}
\end{equation}

so that the predicted trajectory exactly satisfies the
initial condition of \eqref{eq:cts}.

For the uniform knot sequence considered in this work,
let $h$ denote the spacing between consecutive interior
knots. We define the temporal separation between
consecutive Greville abscissae as
\begin{equation}
\Delta_k
:=
\tau_{k+1}^*-\tau_k^*.
\label{eq:greville_spacing}
\end{equation}
For uniformly spaced interior knots, $\Delta_k=h$ for
consecutive interior Greville abscissae, while the clamped
boundary construction produces nonuniform spacing near
the endpoints.

Here, $h$ characterizes the resolution of the B-spline
representation, whereas $\Delta_k$ denotes the temporal
separation between consecutive Greville abscissae.

For a BNO with trained network parameters $\theta$, we
denote the approximation error over the finite horizon
$[0,T]$ by

\begin{equation}
\varepsilon_{\rm BNO}(h,T;\theta)
=
\sup_{x_0\in K}
\left\|
\mathcal{P}(x_0)
-
\widehat{\mathcal{P}}_{\theta,h}(x_0)
\right\|_{L^\infty([0,T])}.
\label{eq:bno_error}
\end{equation}

The dependence on $h$ accounts for the B-spline
representation, while $T$ specifies the prediction
horizon and $\theta$ denotes the trained network
parameters. When these dependencies are not essential
to the discussion, we write simply
$\varepsilon_{\rm BNO}$ to avoid overloading the
notation.

The following approximation result summarizes the
expressive power of B-spline neural operators
\cite{romagnoli_hybrid,wang2025physics}.

\begin{theorem}
\label{thm:universal_approx}
Let $K\subset D$ be compact and let

\[
\mathcal{P}
:
K
\rightarrow
C([0,T],\mathbb{R}^n)
\]

denote the solution operator associated with
\eqref{eq:cts}. For every $\varepsilon>0$, there exist
a B-spline degree $d$, a number of control points
$\ell$, and a neural network generating the
control-point sequence $c(x_0)$ such that

\begin{equation}
\sup_{x_0\in K}
\left\|
\mathcal{P}(x_0)
-
\widehat{\mathcal{P}}(x_0)
\right\|_{L^\infty}
<
\varepsilon,
\label{eq:universal_approx}
\end{equation}

where $\|\cdot\|_{L^\infty}$ denotes the supremum norm
on
$C([0,T],\mathbb{R}^n)$.
\end{theorem}

The tolerance $\varepsilon$ in
Theorem~\ref{thm:universal_approx} is an arbitrary
target accuracy and should be distinguished from
$\varepsilon_{\rm BNO}(h,T;\theta)$, which denotes the
realized approximation error of a particular trained
BNO. The theorem states that, for a fixed finite
horizon $T$, this error can in principle be made
arbitrarily small by selecting a suitable B-spline
representation and a sufficiently expressive neural
network. It does not imply that
$\varepsilon_{\rm BNO}(h,T;\theta)$ decreases as
$h\rightarrow0$ when the trained network parameters
$\theta$ are held fixed. Accordingly, refinement of the
B-spline representation and reduction of the neural
approximation error are treated as distinct
approximation mechanisms in the subsequent analysis.

\revision{The subsequent analysis investigates the relation
between this native control-point representation and
the underlying continuous-time dynamics, providing the
basis for direct spectral and stability analysis of the
trained BNO.}

\section{Problem Statement}

The BNO introduced in the previous section provides a
finite-dimensional representation of the solution of
\eqref{eq:cts} through a sequence of vector-valued
control points
\[
c_1,c_2,\ldots,c_\ell,
\qquad
c_k\in\mathbb{R}^n.
\]

\revision{Unlike samples of the state trajectory, the control
points are coefficients of a B-spline representation
and are not states of the original dynamical system.
Moreover, they are not learned as Koopman observables.
Consequently, their relation to the underlying
continuous-time dynamics is not immediate.}

\revision{The objective of this work is to characterize the
relation between the ordered control-point sequence and
the underlying continuous-time dynamics, and to use
this relation for direct data-driven spectral and
stability analysis of the trained BNO.}

Specifically, the analysis first establishes a
quantitative relation between each B-spline control
point and the underlying continuous-time trajectory.
This relation is then used to derive a dynamical
relation between consecutive control points and the
sampled flow of the original system. On this basis,
the ordered control-point sequence is used directly as
sequential data for DMD and HDMD. Finally, the spectral
information recovered from this representation is
related to qualitative properties of the underlying
system, with particular emphasis on asymptotic
stability.

\section{Dynamical Representation of the Control Points}

\revision{This section establishes the relation between the control-point sequence generated by a trained BNO and the underlying continuous-time dynamics. We first quantify the approximation of the continuous-time trajectory by the control points and then use this result to derive a dynamical relation between consecutive control points and the sampled flow of the original system.}

\subsection{Local Approximation of the Control Points}

\revision{Let
\[
x_k:=x(\tau_k^*),
\]
where $\tau_k^*$ denotes the $k$-th Greville abscissa, and let
$c_k$ denote the corresponding control point generated by the trained
BNO. The sampled trajectory $\{x_k\}$ is introduced here only as an
auxiliary construction to characterize the relation between the
control points and the underlying continuous-time trajectory.}

\begin{lemma}[Control-point approximation]
\label{lem:control_point_error}

Assume the hypotheses of
Lemma~\ref{lem:control_point_proximity} hold and let
\begin{equation}
\hat{x}(t)
=
\sum_{i=1}^{\ell} B_{i,d}(t)c_i
\label{eq:trajectory_drop}
\end{equation}
denote the trajectory generated by the trained
BNO. Then, for every control point,
\begin{equation}
c_k
=
x(\tau_k^*)+\delta_k,
\label{eq:control_point_error}
\end{equation}
where
\begin{equation}
\|\delta_k\|
\leq
\widetilde C_d h^2
\|\ddot{\hat{x}}\|_{L^\infty(J_k)}
+
\varepsilon_{\rm BNO}(h,T;\theta),
\label{eq:delta_bound}
\end{equation}
with
\[
\widetilde C_d := C_d(d-1)^2,
\]
$J_k=[\tau_{k+1},\tau_{k+d}]$, and
$\widetilde C_d>0$ depending only on the spline degree
$d$.

\end{lemma}

\begin{proof}

By adding and subtracting the BNO-generated trajectory
$\hat{x}(\tau_k^*)$, the approximation error satisfies
\begin{align}
\|\delta_k\|
&=
\|c_k-x(\tau_k^*)\| \nonumber\\
&=
\|c_k-\hat{x}(\tau_k^*)
+\hat{x}(\tau_k^*)-x(\tau_k^*)\| \nonumber\\
&\leq
\|c_k-\hat{x}(\tau_k^*)\|
+
\|\hat{x}(\tau_k^*)-x(\tau_k^*)\|.
\label{eq:delta_triangle}
\end{align}

By Lemma~\ref{lem:control_point_proximity} and the
uniform knot spacing,
\begin{equation}
\|c_k-\hat{x}(\tau_k^*)\|
\leq
\widetilde C_d h^2
\|\ddot{\hat{x}}\|_{L^\infty(J_k)}.
\label{eq:local_spline_bound}
\end{equation}

Moreover, by definition of the BNO approximation error,
\begin{equation}
\|\hat{x}(\tau_k^*)-x(\tau_k^*)\|
\leq
\|\hat{x}-x\|_{L^\infty([0,T])}
\leq
\varepsilon_{\rm BNO}(h,T;\theta).
\label{eq:bno_error_bound}
\end{equation}

Combining \eqref{eq:delta_triangle},
\eqref{eq:local_spline_bound}, and
\eqref{eq:bno_error_bound} gives
\eqref{eq:delta_bound}.
\end{proof}

Equation~\eqref{eq:control_point_error} provides the
fundamental link between the control points and the
continuous-time dynamics. The bound contains two
contributions: a local B-spline approximation term,
governed by the knot spacing and the curvature of the
BNO-generated trajectory, and the approximation error
$\varepsilon_{\rm BNO}(h,T;\theta)$ of the trained
neural operator.

It is important to emphasize that
\eqref{eq:control_point_error} does not imply that the control
points should be interpreted, or selected, as trajectory samples. Rather, the lemma quantifies the discrepancy between the B-spline coefficients and the values of the underlying trajectory at the Greville abscissae. Thus, it does not imply that replacing the control points by trajectory samples would yield a smaller approximation error. For finite knot spacing, the control points remain the coefficients of the B-spline representation \cite{unser1992polynomial,unser1993b}, where B-spline representations are constructed from conventionally sampled signals. Here, the trajectory values at the Greville abscissae are introduced only as auxiliary quantities to characterize the relation between the learned coefficients and the underlying continuous-time solution.

For a fixed trained BNO, refinement of the B-spline
representation reduces the local spline contribution
in \eqref{eq:delta_bound}, provided that the curvature
of the BNO-generated trajectory remains uniformly bounded.
However, refinement alone does not imply that
$\varepsilon_{\rm BNO}(h,T;\theta)$ vanishes. Therefore,
convergence of the control points to the underlying
trajectory requires both the spline approximation term
and the BNO approximation error to vanish.

Exact convergence is not required for the proposed
analysis. For a finite control-point representation,
the bounded relation established in
Lemma~\ref{lem:control_point_error} provides the basis
for investigating how the ordered control-point sequence
is dynamically related to the sampled continuous-time
flow, as developed in the next subsection.

\subsection{Dynamical Evolution of the Control-Point Representation}

Lemma~\ref{lem:control_point_error} provides a bounded
relation between each control point and the state of the
underlying continuous-time system at the corresponding
Greville abscissa. We now use this relation to characterize
the evolution of consecutive control points.

\begin{proposition}
\label{prop:quasi_markov}

Consider the finite sequence of control points generated by a trained
BNO and let the hypotheses of
Lemma~\ref{lem:control_point_error} hold. Let
\[
    x_k := x(\tau_k^*),
    \qquad
    \Delta_k := \tau_{k+1}^*-\tau_k^*,
\]
so that
\[
    x_{k+1}=\Phi_{\Delta_k}(x_k).
\]
Assume that the control points $c_k$ and the corresponding
states $x_k$ belong to a compact set
$\mathcal K_c\subset D$ on which the sampled-flow maps
$\Phi_{\Delta_k}$ are Lipschitz continuous, with uniformly
bounded Lipschitz constants
\[
L_{\Phi,k}\leq\bar L_\Phi<\infty.
\]

Then the control-point sequence satisfies
\begin{equation}
    c_{k+1}
    =
    \Phi_{\Delta_k}(c_k)+r_k,
    \label{eq:quasi_markov}
\end{equation}
where $r_k$ is a bounded representation residual that satisfies
\begin{equation}
    \|r_k\|
    \leq
    L_{\Phi,k}\|\delta_k\|
    +
    \|\delta_{k+1}\|.
    \label{eq:residual_bound}
\end{equation}
Consequently,
\begin{equation}
    \|r_k\|
    \leq
    (\bar L_\Phi+1)\bar\delta
    =:\bar e,
\end{equation}
where $\bar\delta$ is defined in
\eqref{eq:uniform_delta_bound}.
\end{proposition}
\medskip
\begin{proof} From Lemma~\ref{lem:control_point_error},
\begin{equation}
    c_k=x_k+\delta_k,
    \label{eq:cp_state_relation_prop}
\end{equation}
with
\begin{equation}
    \|\delta_k\|
    \leq
    \widetilde C_d h^2
    \|\ddot{\hat{x}}\|_{L^\infty(J_k)}
    +
    \varepsilon_{\rm BNO}(h,T;\theta).
    \label{eq:delta_prop_bound}
\end{equation}
Similarly,
\begin{equation}
    c_{k+1}=x_{k+1}+\delta_{k+1}.
    \label{eq:cp_state_relation_next}
\end{equation}
Since $x_{k+1}=\Phi_{\Delta_k}(x_k)$ and $x_k=c_k-\delta_k,$
it follows that
\begin{equation}
    c_{k+1}
    =
    \Phi_{\Delta_k}(c_k-\delta_k)
    +
    \delta_{k+1}.
    \label{eq:cp_flow_intermediate}
\end{equation}

Adding and subtracting
$\Phi_{\Delta_k}(c_k)$ gives
\begin{align}
    c_{k+1}
    &=
    \Phi_{\Delta_k}(c_k)
    +
    \Phi_{\Delta_k}(c_k-\delta_k)
    -
    \Phi_{\Delta_k}(c_k)
    +
    \delta_{k+1}
    \nonumber\\
    &=
    \Phi_{\Delta_k}(c_k)+r_k,
\end{align}
where
\begin{equation}
    r_k
    :=
    \Phi_{\Delta_k}(c_k-\delta_k)
    -
    \Phi_{\Delta_k}(c_k)
    +
    \delta_{k+1}.
    \label{eq:representation_residual}
\end{equation}

Because both $c_k$ and
$c_k-\delta_k=x_k$ belong to the compact set
$\mathcal K_c$ on which $\Phi_{\Delta_k}$ is Lipschitz,
we obtain
\begin{align}
    \|r_k\|
    &\leq
    \left\|
    \Phi_{\Delta_k}(c_k-\delta_k)
    -
    \Phi_{\Delta_k}(c_k)
    \right\|
    +
    \|\delta_{k+1}\|
    \nonumber\\
    &\leq
    L_{\Phi,k}\|\delta_k\|
    +
    \|\delta_{k+1}\|.
    \label{eq:residual_intermediate}
\end{align}

Since the control-point sequence is finite, define
\begin{equation}
    \bar\delta
    :=
    \widetilde C_d h^2
    \max_{1\leq k\leq\ell}
    \|\ddot{\hat{x}}\|_{L^\infty(J_k)}
    +
    \varepsilon_{\rm BNO}(h,T;\theta).
    \label{eq:uniform_delta_bound}
\end{equation}
Then
\[
    \|\delta_k\|
    \leq
    \bar\delta,
    \qquad
    k=1,\ldots,\ell.
\]

Using
\[
    L_{\Phi,k}
    \leq
    \bar L_\Phi
\]
in \eqref{eq:residual_intermediate} yields
\begin{equation}
    \|r_k\|
    \leq
    (\bar L_\Phi+1)\bar\delta.
\end{equation}

Therefore, defining
\begin{equation}
    \bar e
    :=
    (\bar L_\Phi+1)\bar\delta,
    \label{eq:uniform_residual_bound}
\end{equation}
gives
\[
    \|r_k\|
    \leq
    \bar e
    <
    \infty,
    \qquad
    k=1,\ldots,\ell-1.
\]

Hence, the ordered control-point sequence satisfies the
bounded dynamical relation
\eqref{eq:quasi_markov}.
\end{proof}

\begin{remark}
\label{rem:representation_residual}

The residual $r_k$ in
\eqref{eq:quasi_markov} is not an exogenous disturbance
acting on the physical system. It arises from expressing
the evolution of the BNO control-point representation
relative to the sampled continuous-time flow.
The decomposition is used only to establish this
dynamical relation; $\Phi_{\Delta_k}$ and $r_k$ are not
identified separately in the subsequent data-driven
analysis.
\end{remark}

The significance of \eqref{eq:quasi_markov} is that it provides a dynamical interpretation of the ordered control-point sequence. Although the control points are B-spline coefficients rather than trajectory samples, Proposition~\ref{prop:quasi_markov} shows that
their evolution can be described through a one-step dynamical
relation with a bounded representation residual. This allows the
control points to be treated directly as sequential dynamical data and provides the basis for applying DMD and HDMD to their complete observed evolution.

\subsection{Connection with Koopman Operator Theory}

In applying DMD and HDMD to the control-point sequence, the objective is not to approximate the sampled flow $\Phi_{\Delta_k}$ appearing in \eqref{eq:quasi_markov}. Rather, the identified operators approximate the complete observed evolution $c_k\mapsto c_{k+1}$, including the effect represented by the bounded residual $r_k$. Thus, $\Phi_{\Delta_k}$ and $r_k$ provide a decomposition used to establish the dynamical relation of Proposition~\ref{prop:quasi_markov}, whereas
DMD and HDMD provide finite-dimensional linear approximations of the control-point evolution as a whole.

The control-point sequence can therefore be used directly as sequential data within the data-driven Koopman framework introduced in Section~II. The control points are not assumed to be Koopman eigenfunctions or to span an exact Koopman-invariant subspace.

For DMD, the data coordinates are chosen as
\begin{equation}
    z_k=c_k,
    \label{eq:dmd}
\end{equation}
and the corresponding finite-dimensional evolution is
approximated by
\begin{equation}
    z_{k+1}\approx A_{\mathrm D}z_k.
    \label{eq:dmd_evolution}
\end{equation}
The spectrum of $A_{\mathrm D}$ therefore characterizes
the linear evolution identified from the complete control-point
sequence.

Alternatively, HDMD enriches the data coordinates using
delayed control points,
\begin{equation}
    z_k=
    \begin{bmatrix}
        c_k^\top &
        c_{k-1}^\top &
        \cdots &
        c_{k-q+1}^\top
    \end{bmatrix}^{\!\top},
    \label{eq:hdmd}
\end{equation}
and identifies
\begin{equation}
    z_{k+1}\approx A_{\mathrm H}z_k.
    \label{eq:hdmd_evolution}
\end{equation}
The delay embedding provides a richer coordinate space
for representing temporal structure that may not be
captured by the instantaneous control point alone.

The resulting spectral information characterizes the
finite-dimensional dynamics identified from the trained
BNO representation. The following section investigates
how this information can be related to stability
properties of the underlying continuous-time system.

\section{Stability Analysis}

The preceding results provide a dynamical connection between
the control-point representation generated by a trained BNO
and the underlying continuous-time system.

We now study how the asymptotic behavior of the continuous-time system is reflected in the control-point representation. We first consider an ideal infinite-horizon setting to establish an equivalence
between convergence of the continuous-time trajectory and
convergence of the associated control-point sequence. The
practical finite-horizon setting is subsequently addressed
through data-driven spectral analysis using DMD and HDMD.

\subsection{Infinite-Horizon Stability Analysis}

Throughout this subsection, we consider the idealized
setting in which the prediction horizon of the B-spline
representation tends to infinity. This assumption is
introduced only to establish the asymptotic relationship
between the control-point representation and the underlying
nonlinear system. The practical finite-horizon setting is
addressed in Section~VI-B through data-driven spectral
analysis using DMD and HDMD.

Over a finite prediction horizon,
Lemma~\ref{lem:control_point_error} quantifies the relation
between the control points and the underlying continuous-time
trajectory. To extend this relation to the ideal
infinite-horizon setting considered here, we introduce the
following asymptotic consistency assumption.

\begin{assumption}
\label{ass:representation_error}
Consider an infinite-horizon B-spline representation
with a locally finite, unbounded uniform knot sequence
of spacing $h$ and Greville abscissae satisfying
\[
\tau_k^*\rightarrow\infty
\qquad\text{as}\qquad
k\rightarrow\infty.
\]
Assume that there exists a constant bias
$b\in\mathbb{R}^n$ such that
\[
\sup_{t\in J_k}
\left\|
\hat{x}(t;x_0)-x(t;x_0)-b
\right\|
\rightarrow0
\qquad\text{as}\qquad
k\rightarrow\infty.
\]
Moreover, assume that the B-spline representation becomes
asymptotically flat in the sense that
\[
\left\|
\ddot{\hat{x}}
\right\|_{L^\infty(J_k)}
\rightarrow0
\qquad\text{as}\qquad
k\rightarrow\infty.
\]
\end{assumption}
\medskip
\begin{theorem}
\label{thm:cp_stability}
Let $x_e$ be an equilibrium point of \eqref{eq:cts} and
let $\{c_k\}_{k=1}^{\infty}$ denote the corresponding
infinite-horizon control-point sequence. Under
Assumption~\ref{ass:representation_error}, define
\[
\hat{x}_e:=x_e+b.
\]
Then
\[
\lim_{t\rightarrow\infty}x(t;x_0)=x_e
\quad\Longleftrightarrow\quad
\lim_{k\rightarrow\infty}
\|c_k-\hat{x}_e\|=0.
\]
\end{theorem}
\medskip
\revision{\begin{proof}
We prove the two implications separately.
\medskip
\noindent
\emph{Forward implication.}
Suppose that
\[
x(t;x_0)\longrightarrow x_e
\qquad\text{as}\qquad
t\longrightarrow\infty.
\]
From the local B-spline approximation result,
\[
\|c_k-\hat{x}(\tau_k^*;x_0)\|
\le
\widetilde C_d h^2
\|\ddot{\hat{x}}\|_{L^\infty(J_k)}.
\]
Using
\[
\hat{x}_e=x_e+b,
\]
we write
\begin{align}
c_k-\hat{x}_e
&=
c_k-\hat{x}(\tau_k^*;x_0)
\nonumber\\
&\quad+
\hat{x}(\tau_k^*;x_0)
-x(\tau_k^*;x_0)-b
\nonumber\\
&\quad+
x(\tau_k^*;x_0)-x_e .
\end{align}
Hence, by the triangle inequality,
\begin{align}
\|c_k-\hat{x}_e\|
&\le
\|c_k-\hat{x}(\tau_k^*;x_0)\|
\nonumber\\
&\quad+
\|\hat{x}(\tau_k^*;x_0)
-x(\tau_k^*;x_0)-b\|
\nonumber\\
&\quad+
\|x(\tau_k^*;x_0)-x_e\|.
\label{eq:forward_cp_bound}
\end{align}
The first term satisfies
\[
\|c_k-\hat{x}(\tau_k^*;x_0)\|
\le
\widetilde C_d h^2
\|\ddot{\hat{x}}\|_{L^\infty(J_k)}
\longrightarrow0
\]
by Assumption~\ref{ass:representation_error}.
For the second term, since
$\tau_k^*\in J_k$,
\[
\|\hat{x}(\tau_k^*;x_0)
-x(\tau_k^*;x_0)-b\|
\le
\sup_{t\in J_k}
\|\hat{x}(t;x_0)-x(t;x_0)-b\|,
\]
which also converges to zero by
Assumption~\ref{ass:representation_error}.
Finally, because $\tau_k^*\rightarrow\infty$ and $x(t;x_0)\rightarrow x_e,$ we have
\[
x(\tau_k^*;x_0)\rightarrow x_e.
\]
Therefore, every term on the right-hand side of
\eqref{eq:forward_cp_bound} converges to zero, and hence
\[
c_k\longrightarrow\hat{x}_e.
\]
\medskip
\noindent
\emph{Reverse implication.}
Suppose now that
\[
c_k\longrightarrow\hat{x}_e.
\]
Consider the infinite-horizon B-spline representation
\[
\hat{x}(t)
=
\sum_{i=1}^{\infty}
B_{i,d}(t)c_i.
\]
Because the B-spline basis functions have compact support,
the sum is locally finite. Define the active index set
\[
\mathcal I(t)
=
\{i\ge1:B_{i,d}(t)\neq0\}.
\]
Since the basis functions are nonnegative and form a
partition of unity,
\[
B_{i,d}(t)\ge0,
\qquad
\sum_{i\in\mathcal I(t)}
B_{i,d}(t)=1.
\]
It follows that
\begin{align}
\hat{x}(t)-\hat{x}_e
&=
\sum_{i\in\mathcal I(t)}
B_{i,d}(t)(c_i-\hat{x}_e),
\end{align}
and therefore
\begin{align}
\|\hat{x}(t)-\hat{x}_e\|
&\le
\sum_{i\in\mathcal I(t)}
B_{i,d}(t)
\|c_i-\hat{x}_e\|
\nonumber\\
&\le
\max_{i\in\mathcal I(t)}
\|c_i-\hat{x}_e\|.
\label{eq:cp_to_spline_bound}
\end{align}
Since the knot sequence is locally finite and unbounded, the active basis functions shift to increasingly large indices as $t\rightarrow\infty$, namely,
\[
\min\mathcal I(t)\rightarrow\infty.
\]
Because
\[
c_k\rightarrow\hat{x}_e,
\]
for every $\epsilon>0$ there exists $N_\epsilon$ such that
\[
\|c_i-\hat{x}_e\|<\epsilon
\qquad
\forall\, i\ge N_\epsilon.
\]
For sufficiently large $t$, every active index satisfies $i\ge N_\epsilon$, and hence
\[
\|\hat{x}(t)-\hat{x}_e\|<\epsilon.
\]
Therefore,
\[
\hat{x}(t)\longrightarrow\hat{x}_e.
\]
Using
\[
\hat{x}_e=x_e+b,
\]
we write
\begin{align}
x(t;x_0)-x_e
&=
x(t;x_0)+b-\hat{x}_e
\nonumber\\
&=
x(t;x_0)+b-\hat{x}(t;x_0)
+
\hat{x}(t;x_0)-\hat{x}_e.
\end{align}
Hence,
\begin{align}
\|x(t;x_0)-x_e\|
&\le
\|\hat{x}(t;x_0)-x(t;x_0)-b\|
\nonumber\\
&\quad+
\|\hat{x}(t;x_0)-\hat{x}_e\|.
\label{eq:reverse_stability_bound}
\end{align}
Since the knot sequence is locally finite and unbounded,
$t\rightarrow\infty$ corresponds to $k\rightarrow\infty$
for the local intervals $J_k$. Hence, by
Assumption~\ref{ass:representation_error},
\[
\|\hat{x}(t;x_0)-x(t;x_0)-b\|
\rightarrow0.
\]
The second term in
\eqref{eq:reverse_stability_bound}
also converges to zero because
\[
\hat{x}(t)\rightarrow\hat{x}_e.
\]
Consequently,
\[
x(t;x_0)\longrightarrow x_e.
\]
Thus,
\[
x(t;x_0)\longrightarrow x_e
\quad\Longleftrightarrow\quad
c_k\longrightarrow\hat{x}_e,
\qquad
\hat{x}_e=x_e+b,
\]
which completes the proof.
\end{proof}}
When the asymptotic representation bias vanishes, $b=0$, the
control points converge directly to the physical equilibrium $x_e$.



\begin{corollary}
\label{cor:practical_cp_convergence}
Let $x_e$ be an asymptotically stable equilibrium of
\eqref{eq:cts}, and let $x(t;x_0)$ be a trajectory in its region of attraction. Suppose that
\[
\tau_k^*\rightarrow\infty,
\]
and that the BNO representation satisfies
\[
\limsup_{k\rightarrow\infty}
\sup_{t\in J_k}
\|\hat{x}(t;x_0)-x(t;x_0)\|
\le
\varepsilon_\infty
\]
for some $\varepsilon_\infty\ge0$.
Assume also that
\[
\limsup_{k\rightarrow\infty}
\|\ddot{\hat{x}}\|_{L^\infty(J_k)}
\le
M_\infty
\]
for some $M_\infty<\infty$.
Then
\[
\limsup_{k\rightarrow\infty}
\|c_k-x_e\|
\le
\varepsilon_\infty
+
\widetilde C_d h^2 M_\infty.
\]
Hence, the control-point sequence asymptotically remains in a neighborhood of the equilibrium whose size depends on the BNO approximation error and the local B-spline approximation error.
\end{corollary}
\revision{\begin{proof}
From the local B-spline approximation result,
\[
\|c_k-\hat{x}(\tau_k^*;x_0)\|
\le
\widetilde C_d h^2
\|\ddot{\hat{x}}\|_{L^\infty(J_k)}.
\]
Therefore,
\begin{align}
\|c_k-x_e\|
&\le
\|c_k-\hat{x}(\tau_k^*;x_0)\|
\nonumber\\
&\quad+
\|\hat{x}(\tau_k^*;x_0)
-x(\tau_k^*;x_0)\|
\nonumber\\
&\quad+
\|x(\tau_k^*;x_0)-x_e\|.
\end{align}
Since $x_e$ is asymptotically stable and
$x_0$ belongs to its region of attraction,
\[
x(t;x_0)\rightarrow x_e.
\]
Moreover, since $\tau_k^*\rightarrow\infty$,
\[
\|x(\tau_k^*;x_0)-x_e\|
\rightarrow0.
\]
Taking the limit superior and using the assumed bounds on the BNO approximation error and spline curvature gives
\[
\limsup_{k\rightarrow\infty}
\|c_k-x_e\|
\le
\varepsilon_\infty+\widetilde C_dh^2M_\infty.
\]
\end{proof}}

In the ideal infinite-horizon setting, these results show that an asymptotically consistent BNO preserves the equilibrium-convergence behavior of the underlying continuous-time system directly in its native control-point representation. In particular, in the unbiased case, convergence of the continuous-time trajectory to an equilibrium $x_e$ is equivalent to convergence of the corresponding control-point sequence to
the same equilibrium,
\[
x(t;x_0)\rightarrow x_e
\quad\Longleftrightarrow\quad
c_k\rightarrow x_e.
\]

\subsection{Finite-Horizon Stability Assessment}
\label{subsec:finite_horizon}

Theorem~\ref{thm:cp_stability} relates convergence of the
continuous-time trajectories to convergence of the corresponding
control-point sequences. In particular, for $x_0\in K$,
\[
z_k(x_0):=c_k(x_0)-\hat{x}_e\rightarrow0
\quad\Longleftrightarrow\quad
x(t;x_0)\rightarrow x_e .
\]
A trained BNO provides only finite control-point sequences over
$I=[0,T]$. We therefore identify a common finite-dimensional
operator $A_N$ from sequences generated for initial conditions in
$K$ and use its spectral properties to assess the dynamics encoded
by the BNO. DMD uses $o_k=z_k$, whereas HDMD uses the corresponding
delay-embedded coordinates.

The information content of the available data is characterized by
the following richness condition.

\begin{assumption}[Finite-data richness]
\label{ass:koopman}
Consider $N_{\rm tr}$ control-point sequences generated from
initial conditions $x_0^{(i)}\in K$, $i=1,\ldots,N_{\rm tr}$.
For each sequence, let $o_j^{(i)}=z_j^{(i)}$ for DMD and
\[
o_j^{(i)}
=
z_j^{{\rm H},(i)}
=
\begin{bmatrix}
z_j^{(i)\top} &
z_{j-1}^{(i)\top} &
\cdots &
z_{j-q+1}^{(i)\top}
\end{bmatrix}^{\!\top}
\]
for HDMD. Assume
\begin{equation}
G_N:=
\sum_{i=1}^{N_{\rm tr}}
\sum_{j\in\mathcal I_i}
o_j^{(i)}o_j^{(i)\top}
\succeq d_\theta I,
\qquad
d_\theta=\lambda_{\min}(G_N)>0,
\label{eq:richness_condition}
\end{equation}
where $\mathcal I_i$ contains the valid identification indices for
the $i$th sequence.
\end{assumption}
A positive $d_\theta$ ensures that the aggregated control-point data
span the selected identification space, whereas small values indicate
weak excitation or ill-conditioning. Similar finite-data rank
conditions are used in Koopman identification
\cite{mazouchi2023finite}. Under this condition, the aggregated data
are used to identify a common operator $A_N$ describing the
control-point dynamics over the considered set $K$.

\begin{proposition}
\label{prop:finite_horizon}
Suppose Assumption~\ref{ass:koopman} holds and let $A_N$ denote the common DMD or HDMD operator identified from the aggregated control-point data. If the identified dynamics
\begin{equation}
    o_{k+1}=A_No_k
    \label{eq:identified_cp_dynamics}
\end{equation}
represent the corresponding control-point evolution and
$\rho(A_N)<1$, then
\[
    o_k(x_0)\rightarrow0,
    \qquad x_0\in K.
\]
Consequently,
\[
    c_k(x_0)\rightarrow\hat{x}_e,
    \qquad x_0\in K,
\]
and, under the conditions of Theorem~\ref{thm:cp_stability},
\[
    x(t;x_0)\rightarrow x_e,
    \qquad x_0\in K.
\]
\end{proposition}

\begin{proof}
Since $\rho(A_N)<1$, the identified dynamics in
\eqref{eq:identified_cp_dynamics} are Schur stable and therefore
$o_k(x_0)\rightarrow0$. For DMD, $o_k=z_k$, while for HDMD,
$z_k$ is a component of $o_k$. Hence
$c_k(x_0)\rightarrow\hat{x}_e$, and the conclusion follows from
Theorem~\ref{thm:cp_stability}.
\end{proof}

Proposition~\ref{prop:finite_horizon} provides the connection between
the spectrum of the identified control-point dynamics and the
continuous-time convergence characterized by
Theorem~\ref{thm:cp_stability}. In particular, complex eigenvalues
inside the unit circle may produce oscillatory transients, but
$\rho(A_N)<1$ still implies convergence of the identified
control-point dynamics to the equilibrium representation. Since
$A_N$ is obtained from finite BNO data, the accuracy of this
finite-dimensional representation must be assessed together with
the spectral radius; in the experiments, this is evaluated through
the reconstruction error.

\begin{remark}
\label{rem:spectral_interpretation}
The stability interpretation of
Proposition~\ref{prop:finite_horizon} applies when the BNO
control-point representation is consistent with convergence toward an
equilibrium and the identified dynamics adequately represent its
evolution. For periodic or chaotic behavior, however, the control-point
evolution may require richer finite-dimensional coordinates to be
adequately represented, and dynamically relevant modes may not be
captured by the identified DMD or HDMD operator. Consequently, the
spectral radius alone is not sufficient to characterize such
non-equilibrium dynamics and should be interpreted together with the
observed behavior of the BNO control-point representation and the
richness of the identification data.
\end{remark}

\section{Experimental Validation}
\label{sec:experiments}

We evaluate the proposed analysis on an LTI system, the Van der Pol
(VdP) oscillator, and the Lorenz system \cite{brunton2022data}. For each system, a BNO is
trained using $\ell\in\{25,50,100,150,200\}$ control points with cubic
B-splines ($d=3$) and clamped uniform knot vectors. The network is
trained end-to-end through the continuous-time trajectory reconstruction,
with the loss evaluated against numerically integrated reference
trajectories~\cite{romagnoli_hybrid}. The first control point is fixed
by the initial condition, while the network predicts the remaining
coefficients.

Exact DMD and HDMD are applied directly to the centered control-point
sequences, with HDMD embedding dimension $q=2n+1$
\cite{deyle2011generalized}. For each experiment, the snapshot matrices
are constructed by aggregating the control-point sequences from 20
trajectories, and a common DMD or HDMD operator is identified from the
resulting data. The richness condition in
Assumption~\ref{ass:koopman} is evaluated through $d_\theta$ in
\eqref{eq:richness_condition}.

\subsection{Systems Under Investigation}

The LTI system has $n=2$ and closed-loop eigenvalues $-0.5$ and
$-1.3$. The VdP oscillator is considered with $\mu=-1$, for which the
origin is locally asymptotically stable, and $\mu=1$, for which the
origin is unstable and trajectories approach a stable limit cycle.
The Lorenz system uses $\sigma=10$ and $\beta=8/3$, with
$\rho=0.5$ for the stable origin, $\rho=14$ near the stable equilibria
$C^\pm$ and the unstable origin, and $\rho=28$ for the chaotic
attractor. All equilibrium cases use $T=10$, while the VdP limit-cycle
and chaotic Lorenz cases use $T=20$ and $T=3$, respectively.

\subsection{Illustration of the Learned B-Spline Representation}

\begin{figure}[htbp]
    \centering
    \includegraphics[width=\columnwidth]{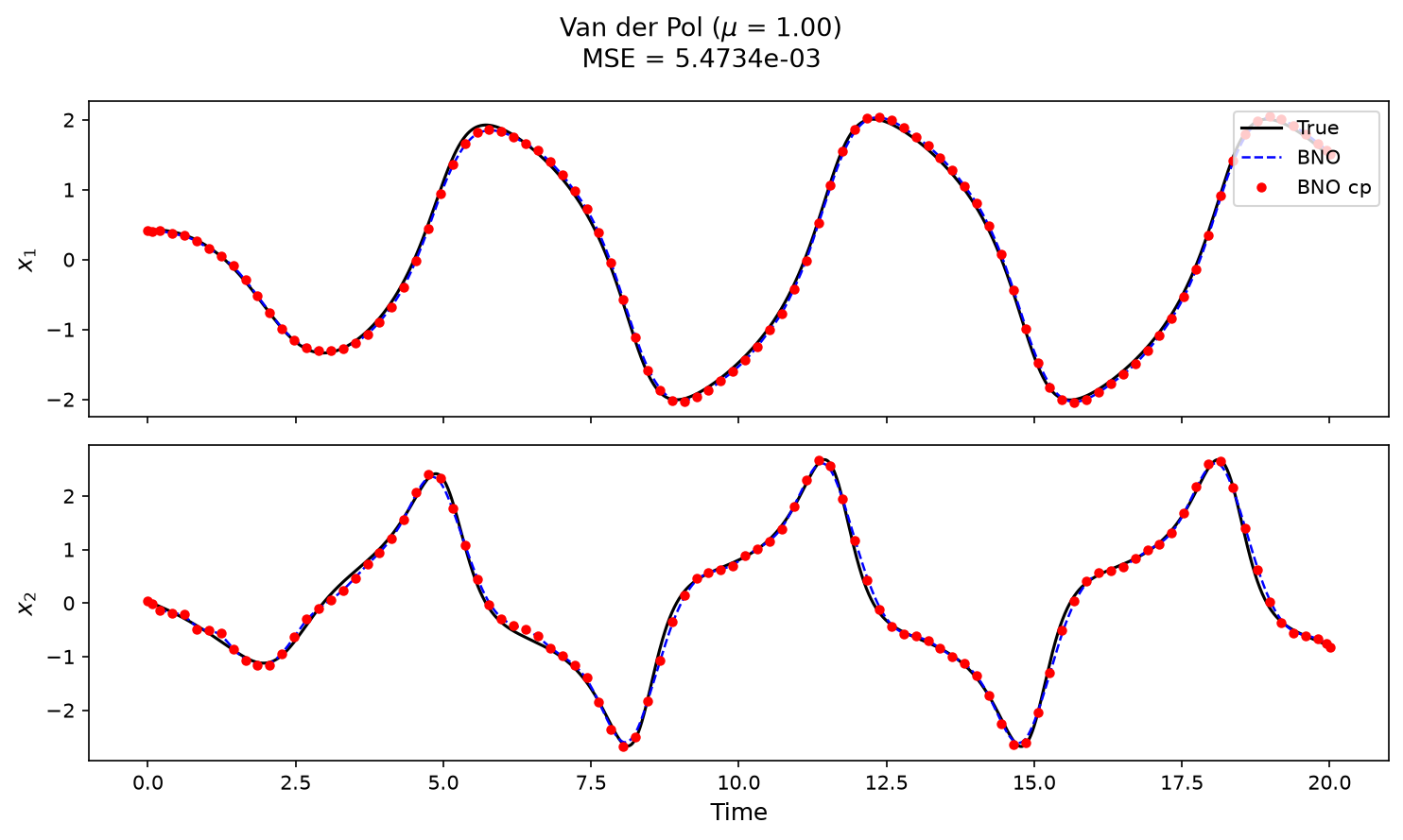}

    \caption{Representative BNO output for the Van der Pol oscillator
    with $\mu=1$. The network predicts the B-spline control points,
    while the continuous trajectory is obtained from the corresponding
    spline expansion.}
    \label{fig:vdp_representation}
\end{figure}

Figure~\ref{fig:vdp_representation} illustrates a representative
BNO output for the VdP oscillator with $\mu=1$. The network predicts
the B-spline control points from which the continuous-time trajectory
is reconstructed. The resulting spline captures both the transient
behavior and the geometry of the stable limit cycle. The control points
are B-spline coefficients rather than trajectory samples and form the
sequence used directly in the subsequent DMD and HDMD analyses.

\subsection{Control-Point Approximation and Connection with the
Continuous-Time Dynamics}

Before analyzing the spectral properties of the control-point
sequences, we numerically examine the relation derived in
Lemma~\ref{lem:control_point_error}. For each trajectory, we define
\begin{align}
e_{\mathrm{CP}}
&=
\max_k
\left\|
c_k-x(\tau_k^*)
\right\|, \\
e_{\mathrm{spl}}
&=
\max_k
\left\|
c_k-\hat{x}(\tau_k^*)
\right\|, \\
e_{\mathrm{BNO}}
&=
\max_t
\left\|
\hat{x}(t)-x(t)
\right\|.
\end{align}
Here, $e_{\mathrm{CP}}$ is the discrepancy characterized in
Lemma~\ref{lem:control_point_error}, $e_{\mathrm{spl}}$ isolates the
coefficient-to-spline discrepancy, and $e_{\mathrm{BNO}}$ measures
the BNO trajectory-approximation error. Values are averaged over the
same 20 trajectories used for the spectral analysis.

\revision{Table~\ref{tab:cp_approximation} reports these quantities at the
coarsest and finest control-point resolutions for the systems with
asymptotically stable equilibria. The corresponding knot spacing is
$h=T/(\ell-d)$.}

\begin{table}[htbp]
\centering
\caption{Control-point approximation errors for the asymptotically
stable systems at the coarsest and finest B-spline resolutions
(mean over 20 trajectories).}
\label{tab:cp_approximation}
\footnotesize
\setlength{\tabcolsep}{3.0pt}
\begin{tabular}{lccccc}
\hline
\textbf{System} & $\bm{\ell}$ & $\bm{h}$ &
$e_{\mathrm{CP}}$ & $e_{\mathrm{spl}}$ &
$e_{\mathrm{BNO}}$ \\
\hline
LTI
& 25  & 0.455 & $6.50\text{e-}2$ & $6.21\text{e-}2$ & $4.58\text{e-}3$ \\
& 200 & 0.051 & $2.02\text{e-}3$ & $1.62\text{e-}3$ & $1.17\text{e-}3$ \\

VdP $\mu=-1$
& 25  & 0.455 & $3.12\text{e-}2$ & $3.10\text{e-}2$ & $2.09\text{e-}3$ \\
& 200 & 0.051 & $4.74\text{e-}3$ & $1.06\text{e-}3$ & $4.31\text{e-}3$ \\

Lor.\ $\rho=0.5$
& 25  & 0.455 & $4.23\text{e-}1$ & $4.56\text{e-}1$ & $1.47\text{e-}1$ \\
& 200 & 0.051 & $4.35\text{e-}2$ & $3.98\text{e-}2$ & $1.20\text{e-}2$ \\

Lor.\ $\rho=14$, $C^+$
& 25  & 0.455 & $7.08$ & $6.27$ & $1.86$ \\
& 200 & 0.051 & $1.32\text{e-}1$ & $1.11\text{e-}1$ & $7.26\text{e-}2$ \\

Lor.\ $\rho=14$, $C^-$
& 25  & 0.455 & $6.95$ & $6.15$ & $1.78$ \\
& 200 & 0.051 & $2.02\text{e-}1$ & $1.09\text{e-}1$ & $1.99\text{e-}1$ \\
\hline
\end{tabular}
\end{table}

\revision{The results show that refinement of the B-spline representation
substantially reduces the coefficient-to-spline discrepancy
$e_{\mathrm{spl}}$ for all of the asymptotically stable systems.
The total control-point discrepancy $e_{\mathrm{CP}}$ also decreases
substantially, although it does not necessarily decrease at the same
rate as $e_{\mathrm{spl}}$. This behavior is consistent with
Lemma~\ref{lem:control_point_error}: refinement reduces the local
B-spline contribution, while the total discrepancy also depends on
the approximation error of the trained BNO. In particular, these
results should not be interpreted as establishing an
$O(h^2)$ convergence rate for $e_{\mathrm{CP}}$, since
$e_{\mathrm{BNO}}$ depends on the independently trained BNO at each
resolution.}

These results provide the numerical link between the local
control-point approximation and the subsequent dynamical analysis.
The errors characterized in Lemma~\ref{lem:control_point_error}
enter the representation-residual bound in
\eqref{eq:quasi_markov}, which provides the theoretical basis
for treating the ordered control points as sequential dynamical data.
For trajectories converging to an equilibrium,
Theorem~\ref{thm:cp_stability} relates this behavior to convergence
of the control-point representation. The following experiments
therefore investigate whether DMD and HDMD applied directly to the
centered control-point sequences recover finite-dimensional spectral
behavior consistent with this asymptotic convergence.

\subsection{Effect of Control-Point Resolution on the Spectral Radius}

This experiment examines how the spectral information
recovered directly from the learned control-point
representation reflects the qualitative behavior of the
underlying continuous-time system and how the resulting
spectral estimates depend on the number of control
points~$\ell$.

\begin{figure}[htbp]
    \centering
    \includegraphics[width=\columnwidth]
    {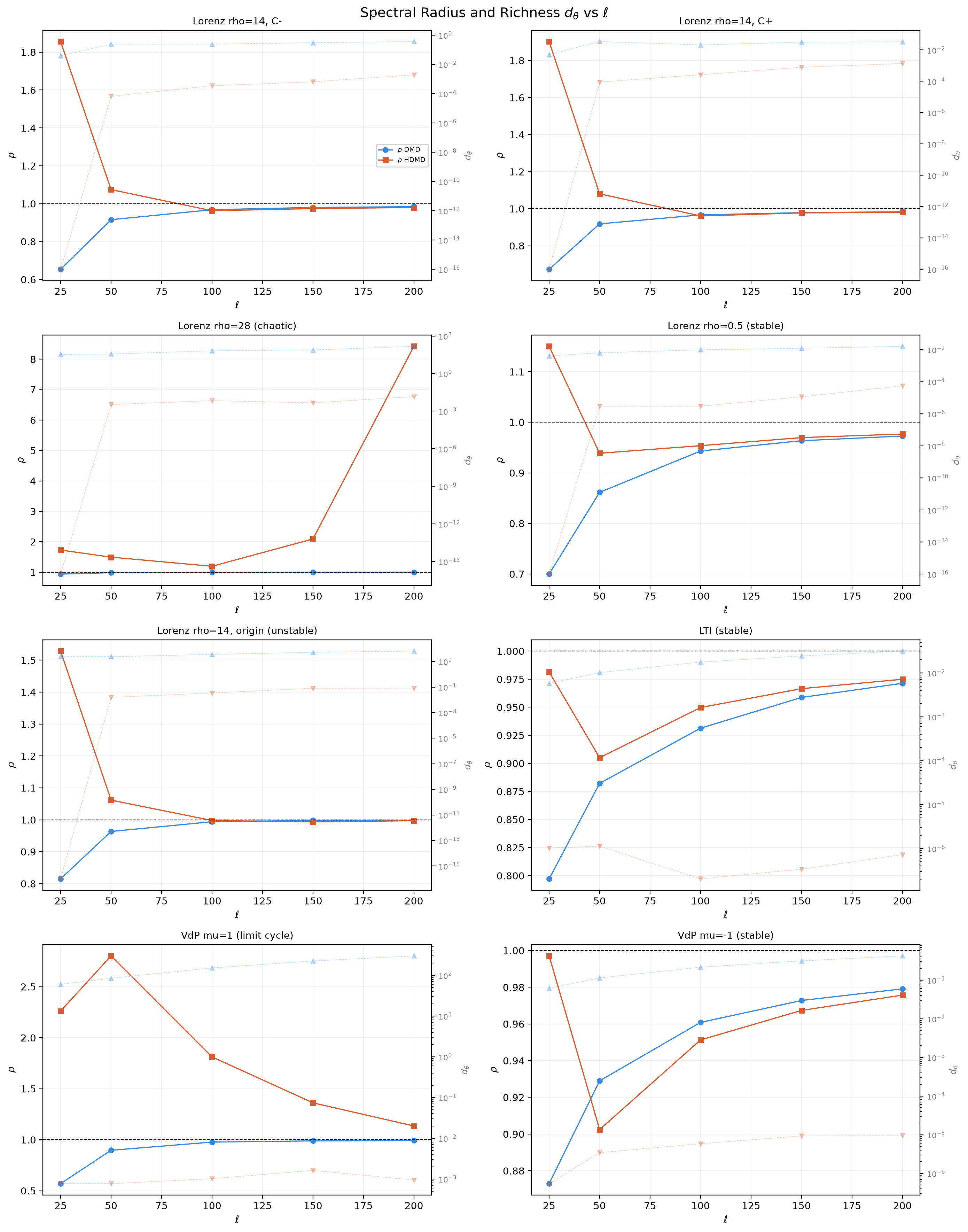}
    \caption{Spectral radius $\rho$ (solid lines, left axis) and
richness constant $d_\theta$ (dashed lines, right axis, logarithmic
scale) as functions of the number of control points $\ell$. The
horizontal dashed line marks $\rho=1$. Values of $d_\theta$ near
machine precision indicate weak excitation or numerical rank
deficiency, making the corresponding spectral estimates potentially
unreliable.}
    \label{fig:rho_vs_ell}
\end{figure}

Figure~\ref{fig:rho_vs_ell} shows the spectral radii of the Exact DMD
and HDMD operators together with the corresponding richness constants
as functions of~$\ell$. The spectral radius is shown on the left axis,
whereas $d_\theta$ is shown on the right axis using a logarithmic scale.

For all asymptotically stable systems, the Exact DMD spectral radius is
strictly smaller than one across the tested resolutions for which the
identification data satisfy the richness condition. At the coarsest
resolution, $\ell=25$, Exact DMD returns
$\rho_{\mathrm{D}}=0.797$ for the LTI system,
$\rho_{\mathrm{D}}=0.873$ for the stable Van der Pol system, and
$\rho_{\mathrm{D}}=0.700$ for the Lorenz system with $\rho=0.5$.
The nonlinear stable equilibria $C^\pm$ of the Lorenz system exhibit
the same qualitative spectral behavior at sufficiently fine
control-point resolutions.

As $\ell$ increases, the spectral radii of the stable systems approach
unity from within. Increasing the resolution reduces the separation
between adjacent Greville abscissae, approximately
$\Delta_\ell=T/(\ell-d)$. For a continuous-time linear mode $\mu_i$,
the corresponding discrete-time eigenvalue is
$\lambda_i(\Delta_\ell)=\exp(\mu_i\Delta_\ell)$; hence,
$\operatorname{Re}(\mu_i)<0$ implies
$|\lambda_i(\Delta_\ell)|\rightarrow1^{-}$ as
$\Delta_\ell\rightarrow0$. The observed trend is therefore consistent
with increasingly fine temporal resolution and does not require the
control points to coincide with trajectory samples.

The Van der Pol limit cycle and the chaotic Lorenz attractor provide
contrasting non-equilibrium cases. As discussed in
Remark~\ref{rem:spectral_interpretation}, the spectral quantities for
these systems are interpreted as finite-horizon characterizations of
the control-point dynamics rather than as certificates of equilibrium
convergence. For the limit cycle, the Exact DMD spectral radius
approaches unity, reaching approximately $0.993$ at $\ell=200$,
consistent with persistent oscillatory behavior rather than convergence
to an equilibrium. For the chaotic Lorenz system, the Exact DMD
spectral radius increases from approximately $0.937$ at $\ell=25$ to
$0.998$ at $\ell=200$.

For the Lorenz system with $\rho=14$ initialized near the unstable
origin, $\rho_{\mathrm{D}}$ approaches unity as $\ell$ increases. The
control-point sequence contains both the initial departure from the
origin and the subsequent convergence toward either $C^+$ or $C^-$.
The resulting finite-horizon DMD operator therefore
characterizes the evolution represented over the complete
observed control-point sequence, including both the departure
from the unstable origin and the subsequent evolution toward
one of the stable equilibria, rather than only the local
behavior near the origin.

\subsection{Interpretation of the Richness Condition}

The effect of finite-data richness is particularly evident for HDMD at
the coarsest resolutions. For the Lorenz system, $n=3$ and $q=7$, so
the embedded identification space has dimension $nq=21$. Although the
delay coordinates are aggregated across the 20 trajectories, the
resulting data can remain weakly excited or nearly linearly dependent.
At $\ell=25$, this is reflected by values of
$d_\theta^{\mathrm H}$ near machine precision, indicating that the
aggregated Gram matrix is numerically ill-conditioned. As $\ell$
increases, the richer delay-coordinate data improve the conditioning
of the identification problem and the HDMD estimates become more
consistent with those obtained using Exact DMD.

A $d_\theta$ sufficiently separated from zero indicates that the
finite-data rank condition is numerically well conditioned, but does
not by itself guarantee an accurate model, since the identification
remains affected by representation error, finite-horizon effects,
conditioning, and model mismatch. This is particularly
important for the limit-cycle and chaotic cases, which satisfy the
richness condition but fall outside the asymptotically stable setting
of Theorem~\ref{thm:cp_stability}. Conversely, a small
$d_\theta^{\mathrm H}$ does not imply that every mode retained after
SVD truncation is unreliable.

\subsection{Trajectory Reconstruction}

\begin{figure}[htbp]
    \centering
    \includegraphics[width=\columnwidth]
    {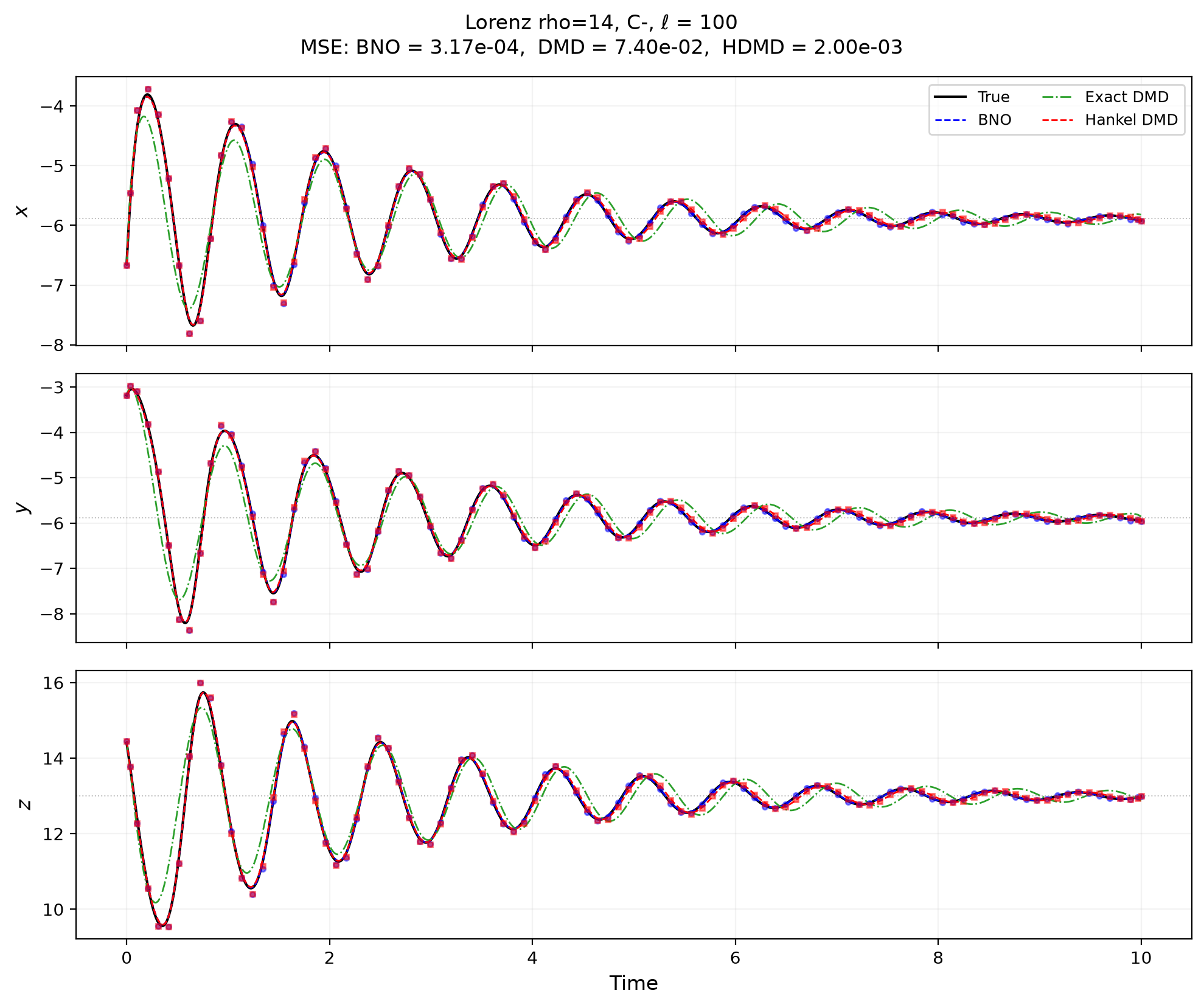}
    \caption{Trajectory comparison for the Lorenz system with
    $\rho=14$ near $C^-$ using $\ell=100$ control points. The BNO
    reconstruction closely tracks the spiraling convergence. The
    HDMD reconstruction more closely reproduces the reference
    trajectory than the Exact DMD reconstruction for this example.}
    \label{fig:trajectory}
\end{figure}

Figure~\ref{fig:trajectory} compares the true solution, BNO
reconstruction, and Exact DMD and HDMD reconstructions for the Lorenz
system with $\rho=14$ near $C^-$ and $\ell=100$.

The BNO, HDMD, and Exact DMD reconstruction errors are
$3.2\times10^{-4}$, $2.0\times10^{-3}$, and
$7.4\times10^{-2}$, respectively. A similar trend occurs for the
stable VdP system at $\ell=100$, where the HDMD and Exact DMD errors
are $8.4\times10^{-6}$ and $3.9\times10^{-4}$.
The improved HDMD reconstruction in these examples reflects its
delay-embedded identification space, but does not imply that HDMD is
universally more reliable, since it requires richer and better
conditioned data.

\subsection{Spectral Analysis}

\begin{figure}[htbp]
    \centering
    \includegraphics[width=\columnwidth]
    {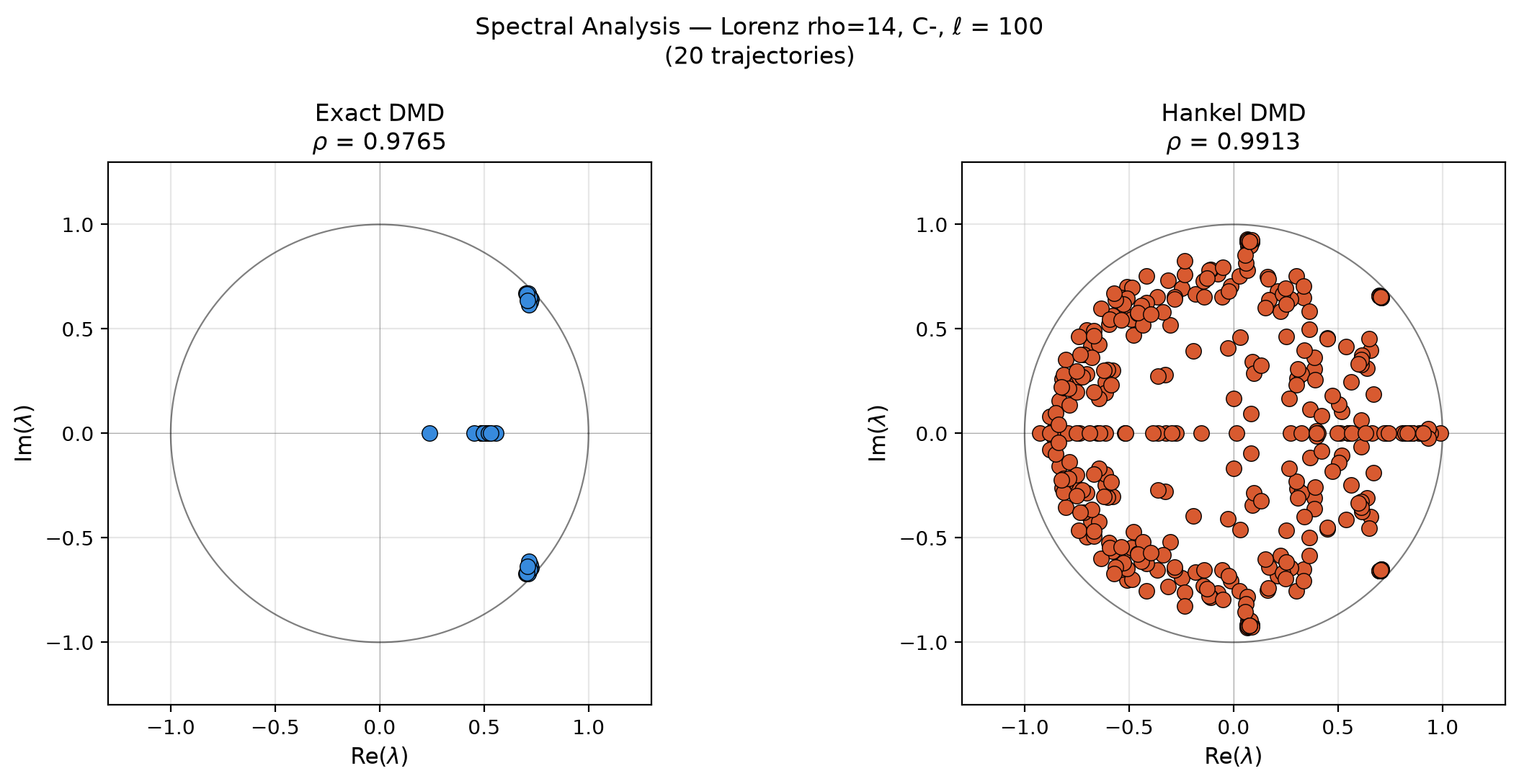}
    \caption{Eigenvalue spectra for the Lorenz system with $\rho=14$
near $C^-$ and $\ell=100$, aggregated over 20 trajectories.
Both Exact DMD and HDMD produce eigenvalues inside the unit disk.
The identified complex-conjugate modes are consistent with the
oscillatory convergence toward the equilibrium.}
    \label{fig:spectra_stable}
\end{figure}

\begin{figure}[htbp]
    \centering
    \includegraphics[width=\columnwidth]
    {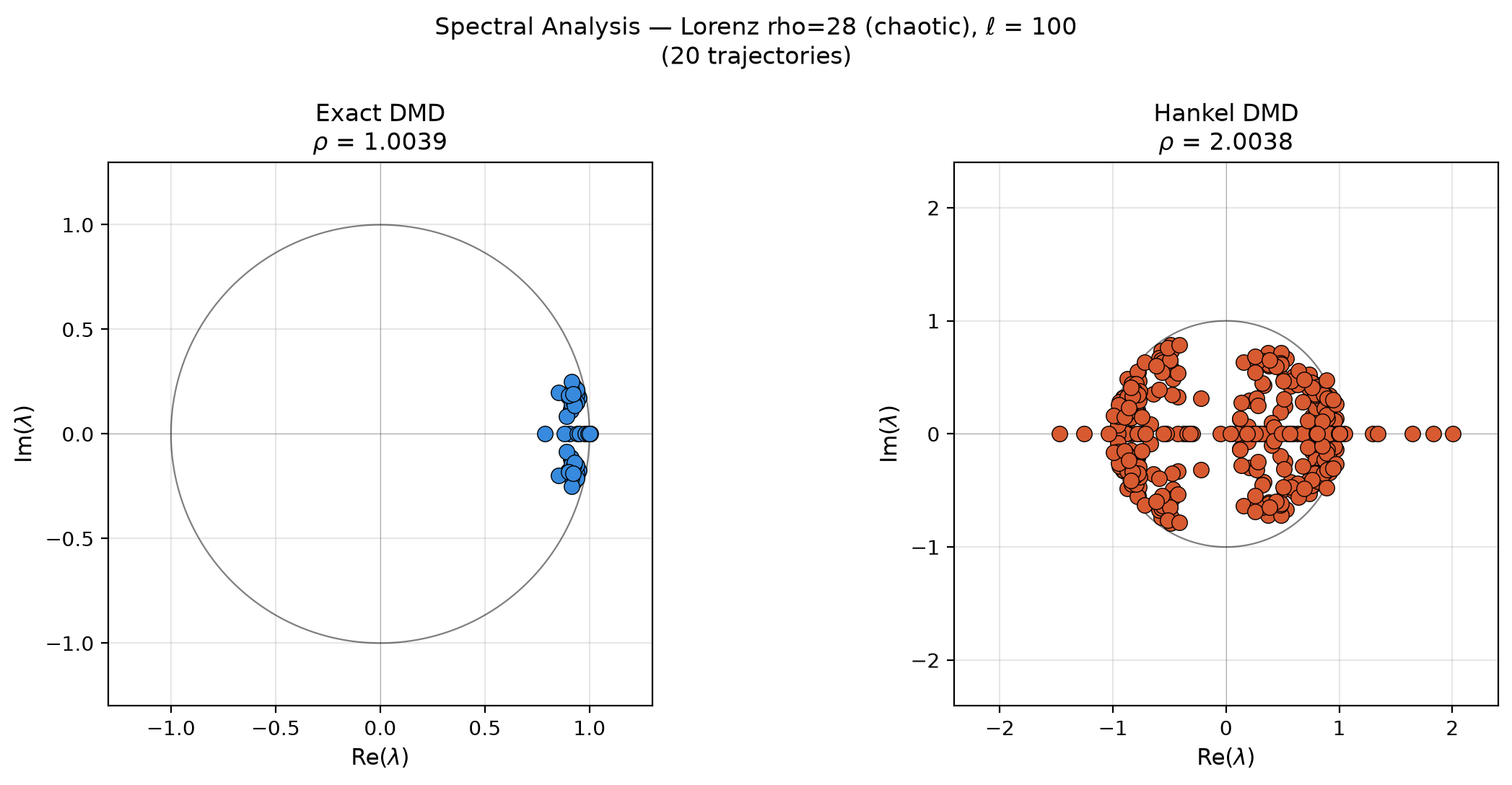}
    \caption{Eigenvalue spectra for the chaotic Lorenz system with
$\rho=28$ and $\ell=100$. The spectra provide finite-dimensional,
finite-horizon characterizations of the control-point representation
and should not be interpreted as asymptotic stability certificates.
HDMD produces a broader distribution of identified modes than
Exact DMD.}
    \label{fig:spectra_chaotic}
\end{figure}

Figure~\ref{fig:spectra_stable} shows representative spectra for the stable Lorenz equilibrium $C^-$, while Figure~\ref{fig:spectra_chaotic} shows the corresponding result for the chaotic regime.

For the stable Lorenz system near $C^-$, both methods yield spectral
radii below unity, $\rho_{\mathrm D}=0.968$ and
$\rho_{\mathrm H}=0.963$, consistent with the converging
control-point behavior and the stability interpretation of
Proposition~\ref{prop:finite_horizon}. For the chaotic case, Exact DMD
gives $\rho_{\mathrm D}=0.993$, whereas HDMD gives
$\rho_{\mathrm H}=1.193$. Although the Exact DMD spectral radius is
below unity, the BNO representation exhibits non-equilibrium chaotic
behavior. Accordingly, as discussed in
Remark~\ref{rem:spectral_interpretation}, the spectrum is interpreted
only as a finite-horizon characterization and not as a certificate of
equilibrium convergence.

\subsection{Summary}

Tables~\ref{tab:stable_results} and~\ref{tab:nonstable_results}
summarize the results at the representative resolution $\ell=100$.
The first table contains the asymptotically stable systems
covered by the stability analysis. The second contains the unstable equilibrium, periodic orbit, and chaotic attractor, which are included to characterize the spectral behavior of the control-point representation outside the main stability result.

\begin{table}[htbp]
\centering
\caption{Spectral analysis of asymptotically stable systems at
$\ell=100$, using control-point data aggregated over 20 trajectories.
The minimum eigenvalue $\lambda_{\min}(G)$ quantifies the finite-data
richness condition in Assumption~\ref{ass:koopman}.}
\label{tab:stable_results}
\footnotesize
\setlength{\tabcolsep}{3.5pt}
\begin{tabular}{lccccc}
\hline
\textbf{System} & \textbf{MSE} &
$\rho_{\mathrm{D}}$ & $\rho_{\mathrm{H}}$ &
$d_\theta^{\mathrm D}$ &
$d_\theta^{\mathrm H}$\\
\hline
Lor.\ $\rho\!=\!0.5$, origin
& $1.8\text{e-}6$ & 0.943 & 0.954
& $9.6\text{e-}3$ & $3.0\text{e-}6$ \\
Lor.\ $\rho\!=\!14$, $C^+$
& $2.4\text{e-}4$ & 0.967 & 0.961
& $2.1\text{e-}2$ & $2.6\text{e-}4$ \\
Lor.\ $\rho\!=\!14$, $C^-$
& $3.2\text{e-}4$ & 0.968 & 0.963
& $2.4\text{e-}1$ & $3.5\text{e-}4$ \\
VdP $\mu\!=\!-1$
& $8.6\text{e-}8$ & 0.961 & 0.951
& $2.2\text{e-}1$ & $5.9\text{e-}6$ \\
\hline
\end{tabular}
\end{table}

\begin{table}[htbp]
\centering
\caption{Spectral analysis of non-equilibrium and unstable systems at
$\ell=100$, using control-point data aggregated over 20 trajectories.
These cases fall outside the asymptotically stable-equilibrium setting
of Theorem~\ref{thm:cp_stability}.}
\label{tab:nonstable_results}
\footnotesize
\setlength{\tabcolsep}{3.5pt}
\begin{tabular}{lccccc}
\hline
\textbf{System} & \textbf{MSE} &
$\rho_{\mathrm{D}}$ & $\rho_{\mathrm{H}}$ &
$d_\theta^{\mathrm D}$ &
$d_\theta^{\mathrm H}$ \\
\hline
Lor.\ $\rho\!=\!14$, origin
& $6.5\text{e-}2$ & 0.994 & 0.998
& $3.6\text{e+}1$ & $3.5\text{e-}2$ \\
Lor.\ $\rho\!=\!28$ (chaotic)
& $8.9$ & 0.993 & 1.193
& $6.5\text{e+}1$ & $6.9\text{e-}3$ \\
VdP $\mu\!=\!1$ (limit cycle)
& $4.3\text{e-}4$ & 0.977 & 1.812
& $1.5\text{e+}2$ & $1.0\text{e-}3$ \\
\hline
\end{tabular}
\end{table}

The stable nonlinear systems yield DMD and HDMD spectral radii below
unity at $\ell=100$, while the unstable, periodic, and chaotic cases
illustrate the different interpretation required outside the setting of
Theorem~\ref{thm:cp_stability}. Overall, the experiments show that the
control-point representation retains sufficient dynamical information
for direct spectral analysis, while $d_\theta$ provides a practical
diagnostic of finite-data richness and conditioning.

\section{Conclusion}

This paper introduced a post-training analysis framework for B-spline
Neural Operators based directly on their control-point representation.
We related the learned control points to the underlying continuous-time
trajectory and characterized their sequential evolution through a bounded
representation residual. This provides a theoretical basis for applying
DMD and HDMD directly to BNO-generated control-point sequences, without
requiring knowledge of the governing differential equations. In the
infinite-horizon setting, convergence of the continuous-time trajectory
to an equilibrium is related to convergence of the control-point sequence
to a learned equilibrium, possibly shifted by approximation bias. Over
finite horizons, DMD and HDMD instead provide a spectral assessment of
the dynamics encoded by the BNO and should not be interpreted as an
unconditional stability certificate for the original nonlinear system.

The numerical results showed spectral behavior consistent with the stable
linear and nonlinear benchmark systems, while also providing
finite-dimensional characterizations of oscillatory and chaotic dynamics.
The identification-data Gram matrix provided an additional diagnostic of
finite-data richness. Although the present work does not establish an
advantage over sample-based spectral identification, it provides a basis
for investigating whether continuous spline representations can alleviate
limitations of snapshot-based methods. Future work will consider
invariant-set computation, region-of-attraction estimation, and Lyapunov
analysis directly from learned control-point representations.

\section*{Generative AI Disclosure}

ChatGPT by OpenAI~\cite{openai_chatgpt} was used as an editorial aid to review the exposition, grammar, and clarity of portions of the manuscript. The authors take full responsibility for the final text.

\section{References}
\bibliographystyle{IEEEtran}
\bibliography{ifacconf}

\end{document}